# A quantitative evaluation of health care system in US, China and Sweden

*Qixin Wang[1], Menghui Li[2], Hualong Zu[3], Mingyi Gao[4], Chenghua Cao[5], Li Charlie Xia[6]*

[1] Department of Mathematics, University of Southern California, Los Angeles, United States of America,
[2] Department of Biomedical Engineering, Peking University, Beijing, PR China,
[3] Department of Electrical Engineering, University of Southern California, Los Angeles, United States of America,
[4] Department of Chemical Engineering and Materials, University of Southern California, Los Angeles, United States of America,
[5] School of Medicine, University of California Irvine, Los Angeles, United States of America,
[6] Department of Biological Sciences, University of Southern California, Los Angeles, United States of America.

**Abstract**

This study is mainly aimed at evaluating the effectiveness of current health care systems of several representative countries and improving that of the US. To achieve these goals, a people-oriented non-linear evaluation model is designed. It comprises one major evaluation metric and four minor metrics. The major metric is constituted by combining possible factors that most significantly determine or affect the life expectancy of people in this country. The four minor metrics evaluate less important aspects of health care systems and are subordinate to the major one. The authors rank some of the health care systems in the world according to the major metric and detect problems in them with the help of minor ones. It is concluded that the health care system of Sweden scores higher than the US and China's system scores lower than that of the US. Especially, the health care system of US lags behind a little bit compared with its economic power. At last, it is reasonable for the American government to optimize the arrangement of funding base on the result of goal programming model.

## Introduction

A satisfactory health system of one country is supposed to provide its residents with effective health care, so that a majority of citizens can enjoy a security and high-quality life, with maximized social equality and minimized total medical expenditure. The complexity of the health care systems makes it difficult to evaluating the health care system by taking into account only a few factors.

The considerations of previous research tend to emphasize the financial efficiency of health care systems. Controlling of the cost has been reported as the key factor of this system [1,2]. However, health care system is slightly different from financial system [3,4]. Medical care is a necessity rather than a commodity for citizens of a country. The quality of medical care of patients is much more important than controlling of cost in health care system [5]. In health care system, health insurance covers most large medical expenses, but there is no institution in a financial system would insurance the high consumption. Thus, we developed a people-oriented comprehensive evaluation system and pay more attention to make sure quality of health care of people. In this evaluation system, the quality of health care to the patient has a higher priority rather than the financial efficiency.

In addition, previous researchers fail to consider the relationship between health care system and medical research institution [1-3]. The health care system is heavily influenced by the development of biomedical research. The investment to the medical research institution can improve the operational efficiency of the health care system, discover new drugs, revise the therapies, and improve the life quality of patients. It is no doubt that increase of investments on medical research has positive impact on health care system in developed country. However, for a developing country, it seems reasonable to pay more attention on other part of health care system such as development medical insurance system, building new hospitals and medical education [6].

Several studies described the qualitative analysis of patient satisfactory of health care and government investment to medical system [7, 8]. However, since there is a complex non-linear relationship between increase of government investment and improvement patient satisfactory of health care system[9,10,11], it is hard to optimize the amount of investment to current health care system; a quantitative analysis model is highly desired. In this study, we are going to conduct a quantitative model to evaluate the health care system of the United States, Sweden, and China. We will also develop a goal programming model to evaluate the best government funding allocation to health care system.

**Definitions and Key Terms**

A *health care system* is the organization and the method by which health care is provided.

The *potential of health care* ($P_{hc}$) of a country shows the power of medical researches and development supported by the government at the present time. It is positively correlated with the size of medical research staff and the quantity of funding from the government.

The health of a citizen is *perfectly ensured*, if his/her income plus aid from the government (plus the financial compensation from medical insurance system if he/she is covered by it) is large enough to prevent and/or cure diseases, ignoring the irreversible damage to health that is beyond the ability of current medical technology.

The *quality index of life* represents the relief from possible diseases or accidentally physical injury based on economic aid offered by medical insurance, and a certain quantity of medical resources provided by current health care system, as well as scientific potential of medicine realized by government-funding researches. This index can approximately imply the quality of life.

The *life expectancy* represents the average life span of a newborn and is an indicator of the overall health of the population of a country.

*Practical effect of medical resources* is the quantity of all categories (medical doctors, nurses, beds) of medical aid that is practically distributed to each citizen in a country on average.

The medical care resources are divided into *essential health care* and *complementary health care resources*. Complementary care is the kind of services that offer holistic benefits that complement or enhance the health care received from the physicians or hospital, and essential health care embraces all the other kinds.

The *matching degree* of a health care system in a country measures whether the system is massive enough to keep up with the development of the country. The health of a citizen is *perfectly ensured*, if his/her income plus aid from the government (plus the financial compensation from medical insurance system if he/she is covered by it) is large enough to prevent and/or cure diseases, ignoring the irreversible damage to health that is beyond the ability of current medical technology.

The *fairness index* represents how well a health care system distributes its resources to everyone who needs it, both rich and poor, urban and rural residents.

The *life index* of a nation is a general and comprehensive figure that describes how much life of high quality is enjoyed by all the citizens in one country. It is positively correlated with *quality index of life* and average *life expectancy*.

*Universal health care* refers to delivery by a combination of public and private systems. In most cases, the law says that everyone must have access to health care. Germany and Sweden, for example, has universal coverage, and social insurance plans cover the majority of people. Symbols are listed in Table 1.

*Assumptions*

People around the world have the same susceptibility to diseases, whichever country they are in.
- Medical personnel and scientific researchers are all competent for their job.
- The per capita GDP of one country can denote how rich and developed the country is.
- Every health care system possesses approximately equal ability of emergency management.
- Every type of disease occurs to people in all countries with the same possibility.
- If a resident is covered by the health care insurance, he/she will be able to afford his/her medical expenditure.
- The investment into scientific medical researches is always effectively used.
- The investment into science researches will pay off (be transformed into applied technology) 25 years later averagely.

*Table 1. Symbols of evaluation model*

| Symbols | Definitions & Descriptions |
|---|---|
| $L_{index}$ | Life index |
| $E_{life}$ | Life expectancy |
| $Q_{life}$ | Life quality index |
| $P_{mr}$ | Practical effect of medical resources |
| $P_{ei}$ | Perfect ensurence index |
| $P_{tech}$ | The current power of medical technology |
| $P_{hc}$ | Potential of health care |
| $R_e$ | Essential health care resources |
| $R_c$ | Complementary health care resources |
| $D_{un}$ | Unnecessary degree |
| $D_{ne}$ | Necessity degree |
| $N_{in}$ | Number of residents who are covered by medical insurance |
| $k_{gov}$ | The proportion of government reimbursement in medical expense |
| $X_{med,i}$ | One's medical expenditure which is submitted to Poisson distribution |
| $X_{inc,i}$ | one's net income which is submitted to normal probability distribution |
| $E_e$ | Average essential expenditure to maintain everyday life |
| $X_{inc,i}$ | one's income which is submitted to normal probability attribution |
| $N_s(t)$ | Number of medical researchers |
| $M_s(t)$ | Quantity of funding going to medical research |
| $\tau$ | Time delay |

**Model Design**

*The Major Evaluation Metric General Analysis*

Since the service object of the health care system is people, we perceive that the evaluation metric should reflect how well the length and quality of people's life are guaranteed by the system through providing health care, which is represented by a general concept called *life index*. Then we decompose *life index* into two parts that are mutually independent: *quality index of life* and *life expectancy*, which measure the quality and quantity of residents' life, respectively. We whereafter keep breaking down *quality index of life* into concrete concepts and simple factors. In so doing, the evaluation system model is concretized and operationalized, because: 1) *life index* is quantified and hence computable; 2) it is easier to search and identify associated sources of data.

As we attach great emphasis on the practical effectiveness of health care systems, the *life index* ($L_{index}$) is the final metric that decides whether a health care system is good or not. According to our definition, we have:

$$L_{index} = Q_{life} \times E_{life}, \quad\quad\quad (1)$$

where

$Q_{life}$ is standardized life quality index, and $E_{life}$ is the average life expectancy of the population in one country.

$E_{life}$ of countries in the world, as a basic and useful kind of data, can be easily found from more than one reliable sources of information, but $Q_{life}$ is comparatively abstract and complicated to measure.

Since it is unreasonable to limit $E_{life}$ to a fixed range, $L_{index}$ is not standardized here.

Obviously, a health care system can help promote $Q_{life}$ in many different ways, but we notice that almost every way is realized through one of the following three channels:

I  every country organizes and provides health care resources to its citizens;
II  medical insurance and government offer economic aid so that the patients have access to medical care service; and
III  the government invests money into medical researches so that we will have more advanced medical technology that can cure the currently incurable diseases and prevent unpredictable diseases in the future.

Using three corresponding variables $P_{mr}$, $P_{ei}$ and $P_{tech}$ to measure the effectiveness of the above three channels, we find that $Q_{life}$ is positively correlated with each one of them. Therefore, it is reasonable to define:

$$Q_{life} = (P_{mr} + P_{ei} + P_{tech})/k_q, \quad\quad\quad (2)$$

where

$P_{mr}$ is the *practical effect of medical resources*,
$P_{ei}$ is *perfect ensurence index*,
$P_{tech}$ is the current power of medical technology (all of them are standardized indexes), and $k_q$ is a coefficient to standardize $Q_{life}$.

To get $Q_{life}$, we have to obtain the value of $P_{mr}$, $P_{ei}$ and $P_{tech}$ one by one.

### Quantify and Calculate $P_{mr}$

Since the medical care resources are divided into essential health care and complementary health care resources, $P_{mr}$ should be broken down into two corresponding parts: the practical effect of essential health care and that of complementary health care. Thus, we get

$$P_{mr} = \prod_{i=1}^{3} \frac{R_{e,i}}{R_{e,i}+k_{e,i}} + \prod_{i=1}^{3} \frac{R_{c,i}}{R_{c,i}+k_{c,i}}, \quad (3)$$

Where $R_1$, $R_2$ and $R_3$ respectively refer to the number of medical doctors, nurses and the beds in hospitals,

$R_e$ is a standardized index that denote the essential health care resources,

$R_c$ is a standardized index that denote the complementary health care resources,

$k_{e,i}$ and $k_{c,i}$ are empirical coefficients, and the relationship among them is expressed by

$$R_{e,i} \times k_{e,i} = R_{c,i} \times k_{c,i}$$

Note that:

when $R_{e,i} \to 0$, we have

$$\frac{R_{e,i}}{R_{e,i}+k_{e,i}} \to \frac{1}{k_{e,i}} R_{e,i},$$

which means the practical effect of medical resources is decided by the quantity of medical resources completely (directly proportional to it);

when $R_{e,i} \to \infty$, we get $\frac{R_{e,i}}{R_{e,i}+k_{e,i}} \to 1$,

which means excessive medical resources contributes little and will cause a great waste;

we multiply the monomial $\frac{R_i}{R_i+k_i}$ ($i$=1, 2, 3)

because the lack of any one of $R_i$ will bring serious difficulty to any health care system.

### Calculate $P_{ei}$

The population of one country can be divided into two categories: those who are covered by medical care insurance ($N_{in}$) and those who are not ($N_{un}$). So the proportion of insured people ($P_{insure}$) is given by

$$P_{insure} = \frac{N_{in}}{population} = \frac{N_{in}}{N_{in}+N_{un}}, \quad\quad (4)$$

The perfect ensurence index ($P_{ei}$) actually measure how many people have their health well ensured through either joining health care insurance or paying by their sufficient income. Thus, it is reasonable to finally define $P_{ei}$ as

$$P_{ei} = 1 - P_{uninsure} \times \frac{total\ medical\ expenditure\ shortage}{total\ medical\ expenditure},$$

$$\quad\quad\quad\quad\quad\quad\quad\quad\quad\quad\quad (5)$$

which in fact is

$$P_{ei} = 1 - (1 - P_{insure}) \times \frac{\sum_{i=1}^{N_{un}} \left[ X_{med,i} \times (1 - k_{gov}) - (X_{inc,i} - E_e) \right] \times I_A}{\sum_{i=1}^{n_{Au}} X_{med,i} \times (1 - k_{gov})},$$

$$I_A = \begin{cases} 1, & i \in A \\ 0, & i \notin A \end{cases},$$

$$A = \left\{ X_{M,i}, X_{inc,i} \mid X_{M,i} \times (1 - k_{gov}) > (X_{inc,i} - E_e) \right\},$$

.......................................... (6)

where

$k_{gov}$ is the proportion of government reimbursement in medical expense,

$E_e$ is the average essential expenditure to maintain everyday life,

$X_{med,i}$ is the medical expenditure of someone in the country, which is submitted to Poisson distribution, and

$X_{inc,i}$ is the net income of someone in the country which is submitted to normal distribution.

### Calculate $P_{tech}$

The current power of medical technology ($P_{tech}$) can be well deduced by the potential of health care ($P_{hc}$) years ago, which means $P_{tech}$ can be estimated as $P_{hc}$ with a time delay ($\tau$), because it takes a period of time to transfer scientific investment into scientific products. Some scholars believe that $\tau = 20$–$30$ years and we make it 25 [6] years here.

Firstly, we calculate $P_{hc}$ based on its definition:

$$P_{hc} = \frac{N_s(t)}{k_N + N_s(t)} \times \frac{M_s(t)}{k_M + M_s(t)} \quad \text{............. (7)}$$

where

$N_s(t)$ is the number of medical researchers, and

$M_s(t)$ is the quantity of funding going to medical research.

Note that both of the two factors, medical researchers and money, can enhance $P_{hc}$, but excessive investment (medical researchers and money) gives only limited effect to $P_{hc}$. This truth supports our idea to define $P_{hc}$ this way.

Secondly, we incorporate $\tau$ into $P_{hc}$ to get $P_{tech}$

$$P_{tech} = \frac{N_s(t-\tau)}{k_N + N_s(t-\tau)} \times \frac{M_s(t-\tau)}{k_M + M_s(t-\tau)} \quad \text{.......... (8)}$$

Medical research plays an important role to improve the math expectation of residents' life in future.

$$E_{life} = E_0 + k_{lt} \times P_{hc} \quad \text{........................ (9)}$$

where $E_0$, $k_{lt}$ are coefficients.

With Eq (1), (2), (3), (4), (6), and (8), $L_{index}$ can be expressed by a complicated equation which involves a series of variables.

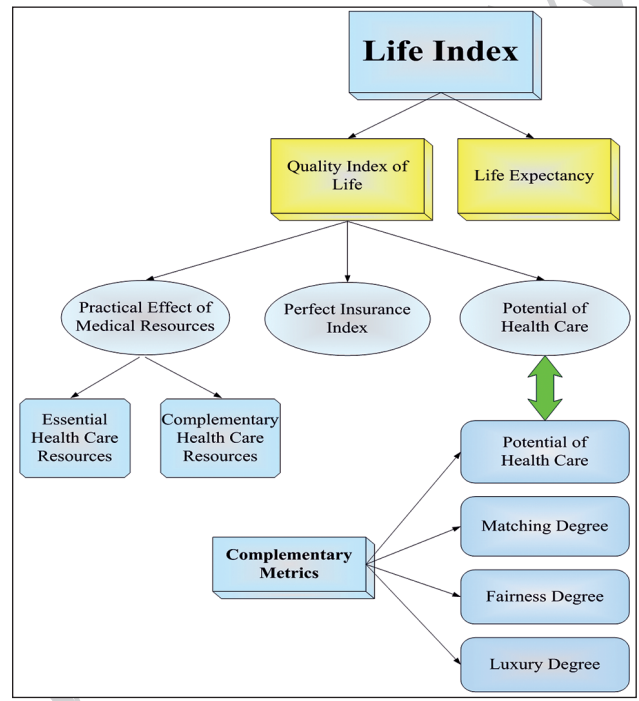

*Figure 1. The diagram showing the organization of evaluation of health care system*

### Subordinate Metrics
### Potential of Health Care ($P_{hc}$)

Eq(7) gives the expression of $P_{hc}$ which is actually a standardized index that predicts the power of medical technology ($P_{tech}$) in the future.

### Matching Degree

To get *matching degree* of each health care system, two factors need to be taken into account: the how well residents' health is ensured and how wealthy the country is. A rich country has the ability to maintain a large scale of health care system that provide abandon health care resources, while a developing country can only afford a smaller and cheaper one. This fact implies that it is harder for a developed country to maintain a matching health care system, because the country has to invest more (money, etc) into its health care system. Therefore, *matching degree* is given by

$$\text{Matching Degree} = \frac{10}{\ln(\text{per capita GDP}) - \ln(L_{index})} \quad (10)$$

where

$R_e$ is the essential health care resources,

$R_c$ is the complementary health care resources, and

*Per capita GDP* is per capital Gross Domestic Product.

### Fairness Degree

The attribution of medical resources cannot possibly be absolutely fair. We tend to believe that wealthy people have more chance to accept medical aid then the poor. Here we compare urban people with those living in rural areas by measuring the quantities of health care resources attributed to them respectively.

$$Fairness\ Degree = \frac{rural\ medical\ resouces}{urban\ medical\ resouces} \quad (11)$$

### Luxury Degree

Considering some parts of a health care system may not play the most essential role or cannot bring immediate benefits to the residents, we define Complementary health care resources and potential of health care to be "unnecessary", while essential health care resources and perfect ensurence index to be "necessary". Thus, if we continue to *define unnecessity degree* ($D_{un}$) and *necessary degree* ($D_{ne}$) as:

$$D_{un} = R_c + P_{hc} \quad\quad\quad (12)$$

$$D_{ne} = R_e + P_{ei} \quad\quad\quad (13)$$

we will arrive at:

$$Luxury\ Index = \frac{D_{un}}{D_{ne} + D_{un}} \quad (14)$$

### Revise the spending plan for US

Although the health care system of US ranked considerably high in the world, is still far from ideal. In this part, we try to revise the previous model to optimize the health care system of US to give more detailed suggestions, as we realize that improving such a complicated system requires further investigation.

Suppose the government has already given it the funding shortage (the health care system of U.S. needs extra 300 billion dollars to push its *matching degree* to as high as Sweden's) and therefore the total budget expands, how shall the health care system spend the extra 300 billion dollars wisely? We argue that a wise spending plan should maximize the *life index*. As we have stated above, *life index* is positively correlated with *quality index of life* and average *life expectancy*. However, *life expectancy* varies very slightly as time elapses, hence we prescribe that an ideal spending plan is the one that maximizes *quality index of life*. Now our aim is to revise the previous model to solve this non-linear programming problem.

We identify nine symbols representing nine major expenditures in Table 2:

*Table 2. Symbols representing major expenditures*

| Symbols | Definition |
|---|---|
| $F_{gov,1}$ | Economic aid to patients |
| $F_{gov,2}$ | Salary of research staff |
| $F_{gov,3}$ | Funding to support medical researches |
| $F_{gov,4}$ | Salary of medical doctors in essential medical source |
| $F_{gov,5}$ | Salary of medical nurses in essential medical source |
| $F_{gov,6}$ | Equipment in essential medical source |
| $F_{gov,7}$ | Salary of medical doctors in complementary medical source |
| $F_{gov,8}$ | Salary of medical nurses in complementary medical source |
| $F_{gov,9}$ | Equipment in complementary medical source |

The Objective function is

$$Max\ L_{index} = (P_{ei} + P_{mr} + P_{hc}) \times (E_0 + k_{lt} \times P_{hc})/k_q$$

$$\quad\quad\quad\quad\quad\quad\quad\quad\quad\quad\quad\quad (15)$$

$$subject\ to\ \sum_{j=1}^{9} F_{gov,j} = F_{total}$$

where

$$P_{ei} = 1 + \left( \frac{F_{income}}{F_{med} - F_{gov,1}} - 1 \right) \times P_{uninsure} EI_A$$

$$P_{hc} = \frac{F_{gov,2}}{k_N S_{salary} + F_{gov,2}} \times \frac{F_{gov,3}}{k_M + F_{gov,3}}$$

$$P_{mr} = \prod_{i=1}^{3} \frac{F_{gov,i+3}}{F_{gov,i+3} + n_{e,1} k_{e,1}} + \prod_{i=1}^{3} \frac{F_{gov,i+6}}{F_{gov,i+6} + n_{c,1} k_{c,1}}$$

The process of deduction shown as following:

$$P_{ei} = 1 - P_{uninsure} \times \frac{\sum_{i=1}^{n_{Au}} (X_{M,i} \times (1 - k_{gov}) - X_{F,i}) \times I_A}{\sum_{i=1}^{n_{Au}} X_{M,i} \times (1 - k_{gov})}$$

$$\Rightarrow P_{ei} = 1 - P_{uninsure} \times \frac{EI_A \times [n_{Au} \times EX_{M,i} \times (1 - k_{gov}) - n_{Au} \times EX_{F,i}]}{n_{Au} \times EX_{M,i} \times (1 - k_{gov})}$$

$$\Rightarrow P_{ei} = 1 - P_{uninsure} \times \frac{EI_A \times [EX_{M,i} \times (1 - k_{gov}) - EX_{F,i}]}{EX_{M,i} \times (1 - k_{gov})}$$

$$\Rightarrow P_{ei} = 1 - P_{uninsure} \times \frac{EI_A \times [EX_{M,i} \times (1 - k_{gov}) - EX_{F,i}]}{EX_{M,i} \times (1 - k_{gov})}$$

$$\Rightarrow P_{ei} = 1 - P_{uninsure} \times EI_A \left[ 1 - \frac{EX_{F,i}}{EX_{M,i} \times (1 - k_{gov})} \right]$$

$$\Rightarrow P_{ei} = 1 - P_{uninsure} EI_A \left[ 1 - \frac{u}{\lambda(1 - \frac{F_{gov,1}}{F_{med}})} \right]$$

$$\Rightarrow P_{ei} = 1 - \left[ P_{uninsure} EI_A - \frac{u P_{uninsure} EI_A}{\lambda \times (1 - \frac{F_{gov,1}}{F_{med}})} \right]$$

$$\Rightarrow P_{ei} = 1 - \left[ P_{uninsure} EI_A - \frac{n_{Au} \times u P_{uninsure} EI_A F_{med}}{n_{Au} \times \lambda \times (F_{med} - F_{gov,1})} \right]$$

$$\Rightarrow P_{ei} = 1 - P_{uninsure} EI_A + \frac{n_{Au} u P_{uninsure} EI_A F_{med}}{n_{Au} \lambda (F_{med} - F_{gov,1})}$$

$$\Rightarrow P_{ei} = 1 - P_{uninsure} EI_A + \frac{F_{income} P_{uninsure} EI_A F_{med}}{n_{Au} \lambda (F_{med} - F_{gov,1})}$$

$$\because F_{med} = n_{Au} \lambda$$

$$\therefore P_{ei} = 1 - P_{uninsure} EI_A + \frac{F_{income} P_{uninsure} EI_A}{F_{med} - F_{gov,1}}$$

$$P_{mr} = \prod_{i=1}^{3} \frac{R_{e,i}}{R_{e,i} + k_{e,i}} + \prod_{i=1}^{3} \frac{R_{c,i}}{R_{c,i} + k_{c,i}}$$

$$\Rightarrow P_{mr} = \prod_{i=1}^{3} \frac{\frac{F_{gov,i+3}}{n_{e,1}}}{\frac{F_{gov,i+3}}{n_{e,1}} + k_{e,1}} + \prod_{i=1}^{3} \frac{\frac{F_{gov,i+6}}{n_{c,1}}}{\frac{F_{gov,i+6}}{n_{c,1}} + k_{c,1}}$$

$$\Rightarrow P_{mr} = \prod_{i=1}^{3} \frac{F_{gov,i+3}}{F_{gov,i+3} + n_{e,1} k_{e,1}} + \prod_{i=1}^{3} \frac{F_{gov,i+6}}{F_{gov,i+6} + n_{c,1} k_{c,1}}$$

$$P_{hc} = \frac{N_s}{k_N + N_s} \times \frac{M_s}{k_M + M_s}$$

$$\Rightarrow P_{hc} = \frac{\frac{F_{gov,2}}{S_{salary}}}{k_N + \frac{F_{gov,2}}{S_{salary}}} \times \frac{F_{gov,3}}{k_M + F_{gov,3}}$$

$$\Rightarrow P_{hc} = \frac{F_{gov,2}}{k_N \times S_{salary} + F_{gov,2}} \times \frac{F_{gov,3}}{k_M + F_{gov,3}}$$

$$\Rightarrow \max L_{index} = (P_{ei} + P_{mr} + P_{hc}) \times (E_0 + k_{lt} \times P_{hc}) / k_q$$

$$\begin{cases} P_{ei} = 1 + \left( \frac{F_{income}}{F_{med} - F_{gov,1}} - 1 \right) \times P_{uninsure} EI_A \\ P_{hc} = \frac{F_{gov,2}}{k_N S_{salary} + F_{gov,2}} \times \frac{F_{gov,3}}{k_M + F_{gov,3}} \\ P_{mr} = \prod_{i=1}^{3} \frac{F_{gov,i+3}}{F_{gov,i+3} + n_{e,1} k_{e,1}} + \prod_{i=1}^{3} \frac{F_{gov,i+6}}{F_{gov,i+6} + n_{c,1} k_{c,1}} \end{cases}$$

subject to $\quad \sum_{j=1}^{9} F_{gov,j} = F_{total}$

## Results and Discussion

### Compare the Effectiveness of Health Care Systems

We had the ability to express $Q_{life}$ with variables that are supported by sufficient data. Figure 2 shows the scores of America, China and Sweden in $P_{mr}$, $P_{ei}$ and $P_{tech}$. Note that the perfect ensurance index of Sweden is 1 (the largest possible value), because, the Sweden has a universal health care system that impose medical insurance to every citizen. Moreover, the power of medical science exceeds that of Sweden, which, we perceive, is an inevitable result of large investment into it in the past several decades.

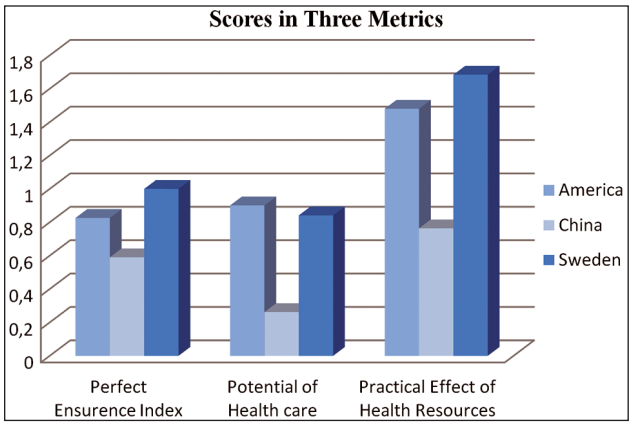

*Figure 2. Three countries' scores of $P_{mr}$, $P_{ei}$ and $P_{tech}$*

Figure 3 provides the comparison of life indexes of US, China and Sweden. Very clearly, the health care system of Sweden is the best among them and the system of China lags behind extremely.

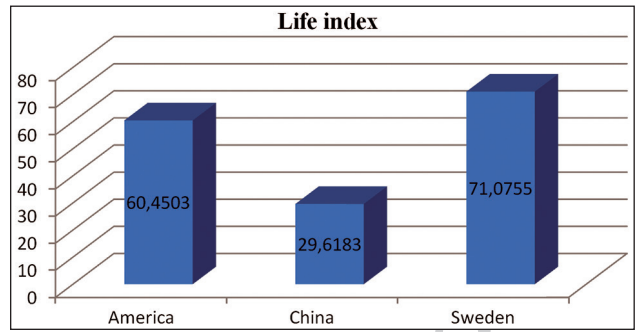

*Figure 3. The current life index of U.S., China and Sweden*

### The analysis of life index of US

Figure 4 shows the life index of US and the change of $P_{mr}$, $P_{ei}$ and $P_{tech}$ from 1990 to 2008; all of the four variables were roughly increasing except for a bit of ramp-down at the end of the 20th century.

The perfect insurance index fluctuates with time, but the general trend is rising, presumably pushed by the development of domestic economy. Since the year 1992, the index rose but dropped drastically in 1998, [14] which is a rather puzzling phenomenon. If we take a look back on history of America, we will find that in the 1993, President Clinton issued a new policy about medical insurance aiming at popularizing medical insurance so that every citizen is covered[16]; in 1998, [15] he declared that this policy was suddenly ceased because of some reason. The data coincide with historic changes amazingly.

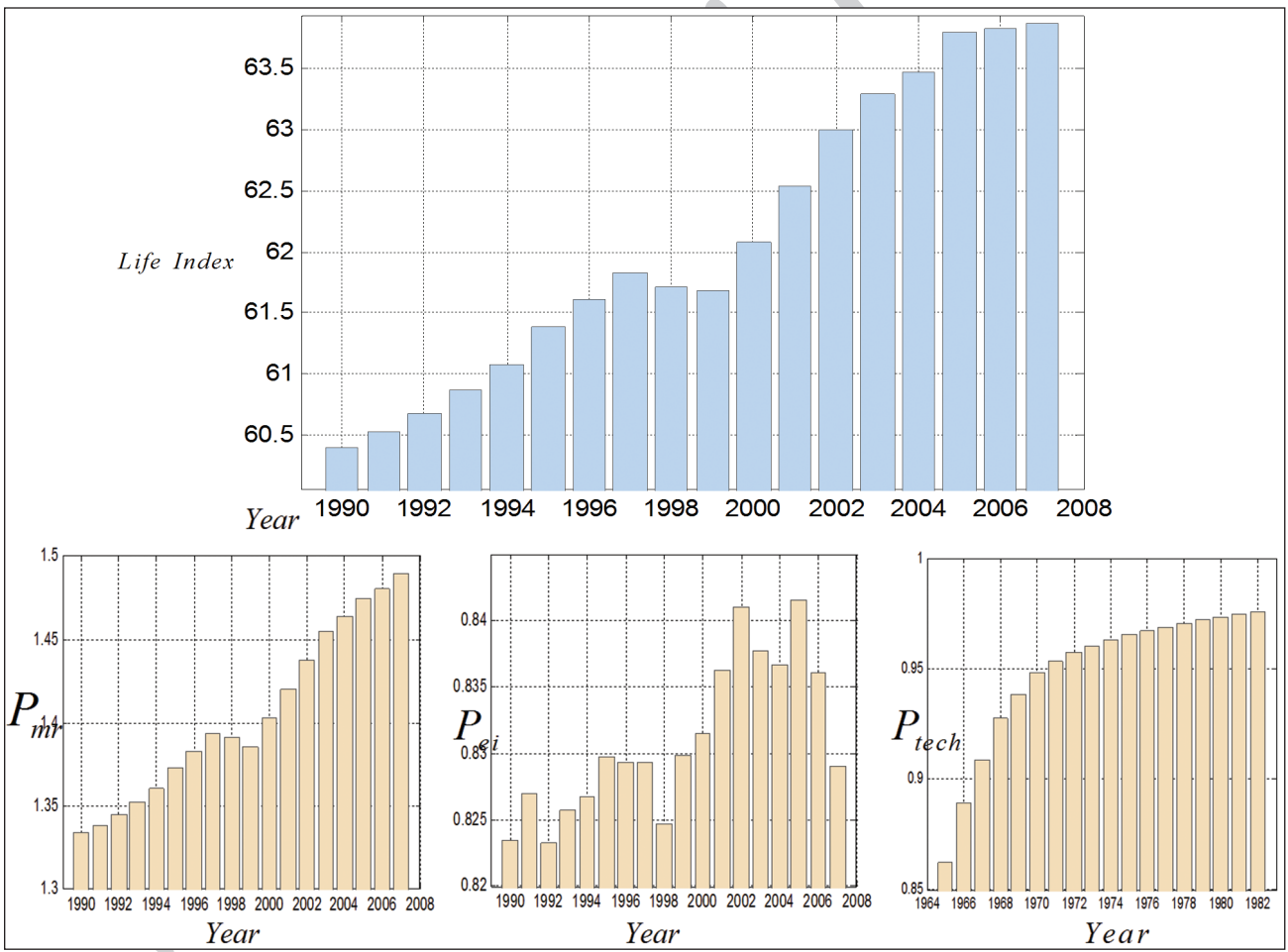

*Figure 4. The history trend of life index, $P_{mr}$, $P_{ei}$ and $P_{tech}$ of US*

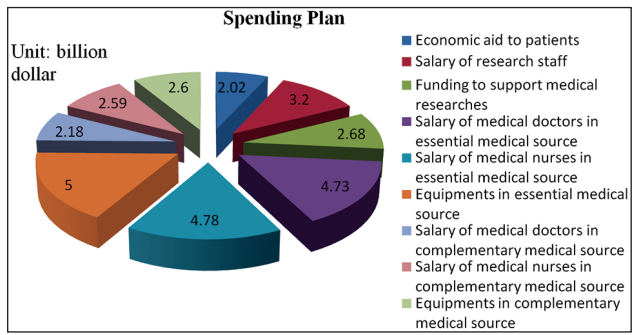

*Figure 5. Revised spending plan of US by greedy algorithm*

### Analyses of Subordinate Metrics

With the most general metric *life index,* we are able to evaluate and rank the health care systems in the world (e.g. the health care system of US is better than that of China but not as good as that of Sweden), but a considerable amount of information is lost or neglected at the same time, which will bring much difficulty in identifying exiting limitations and problems with these health care systems. In order to crack this, we pick out and rearrange some factors to constitute new metrics as complementary metrics (potential of health care, matching degree, fairness degree, & luxury degree). With the help of these complementary metrics, different aspects of one health care system can be evaluated and its limitations become detectable and predictable.

A low *matching degree* may suggest the necessity of investing more money into the health care system so that its scale can be enlarged, while a high one implies that the current health care system is massive enough considering the limited economy scale. Figure 7 implies that the health care system of U.S. should be stronger to match the massive scale of its economy [9]. If we want the matching degree of U.S. to be promoted to 1.66, the government must invest more money into health care system [10]. The *matching degree* implies a slight lack of government investment into health care system [11].

Since it is difficult to obtain all the data to decide their precise quantities, we consider it to be feasible to substitute them with numbers of beds in hospitals in urban and rural areas [12, 13]. The information delivered by Figure 8 is clear: China did a very poor job in health care fairness while that of US could be better.

We consider *luxury index* to be tolerant of subtle conceptual ambiguity, because the slight lack of preciseness in defining concepts may weaken its competence in give an absolute evaluation, but still allows it to serve as a metric to compare different health care systems. (Note that the word 'unnecessary' doesn't mean 'redundant'.)

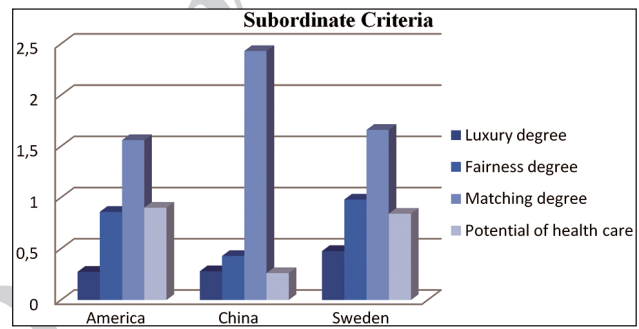

*Figure 6. All the Subordinate metrics*

It is never easy to give a complicated system properly and a precise evaluation [17], as the result is connected with multiple factors that are interwoven with each other [18]. However, if we establish a model that based on reasonable assumptions and tolerate a certain degree of ambiguity, satisfactory result could be achieved [19]. On the other hand, limitations of our model also mainly originate from the assumptions and ambiguity [20-23].

It is well admitted that few things are perfect in the world, whereas we never stop pursuing ideal health care systems, even though it takes a lot of money, manpower, time and energy to improve, because they are our safe guard that relieve our fear of diseases [24-31].

## References

1. Perleth M, Jakubowski E, Busse R, What is 'best practice' in health care? State of the art and perspectives in improving the effectiveness and efficiency of the European health care systems. Health Policy, 2001(56), 235-250

2. Ros CC, Groenewegen PP, Delnoij DM, All rights reserved, or can we just copy? Cost sharing arrangements and characteristics of health care systems, Health Policy, 2000(52), 1-13

3. Abelson J, Miller FA, Giacomini M, What does it mean to trust a health system?: A qualitative study of Canadian health care values, Health Policy 2009(91), 63-70

4. Calnan M, Towards a conceptual framework of lay evaluation of health care, Social Science & Medicine, 1988(27), 927-933

5. Petersen I, Swartz L, Primary health care in the era of HIV/AIDS. Some implications for health systems reform, Social Science & Medicine, 2002(55), 1005-1013

6. Chang C, Zhang Y, Deng D, Xiao Y, A comprehensive evaluation model of health care system, Networking and Information Technology (ICNIT), 2010 International Conference, 535 -539

7. K. Claxton The irrelevance of inference: a decision-making approach to the stochastic evaluation of health care technologies, J Health Econ. 1999(18), PP. 341-64.

8. Yang T, Matthews SA, Understanding the non-stationary associations between distrust of the health care system, health conditions, and self-rated health in the elderly: A geographically weighted regression approach, Health & Place, 2012(18), 576-585

9. Smith PC, Stepan A, Valdmanis V, Verheyen P, Principal-agent problems in health care systems: an international perspective, Health Policy. 1997(41), 37-60

10. Pons-Vigués M, Puigpinós-Riera R, Rodríguez D, Sanmamed M J., Pasarín MI, Pérez G, Borrell C, Casamitjana M, Benet J, Country of origin and prevention of breast cancer: Beliefs, knowledge and barriers, Health & Place, 2012(18), 1270-1281

11. Hollander MJ, Miller JA, Kadlec H, Evaluation of Healthcare Services: Asking the Right Questions to Develop New Policy and Program-Relevant Knowledge for Decision-Making, Healthcare Quarterly, 2010(4), 40-47

12. Oliveira MD et al, Modeling hospital costs to produce evidence for policies that promote equity and efficiency, European Journal of Operational Research, 2008(16), 933-947

13. Rijsbergen MV et al. Managing the overflow of intensive care patients. European Journal of Operational Research, 2008 (16), 988-1010

14. B.X. Qin, Bill Clinton's health care reform. American Research 1994, 7-8

15. Congressional Quarterly, Health Care's Hour, 1993, 19-20

16. Congressional Quarterly, 1993, 2458-2459

17. Wang Q, Liu Y, Mo L, The evaluation and prediction of the effect of AIDS therapy, Proceeding of IEEE/ICME International Conference, 2007, 1591- 1596

18. Wang Q, Liu Y, Pan X, Atmosphere pollutants and mortality rate of respiratory diseases in Beijing, Science of the Total Environment, 2008(391), 143-148

19. Wang Q, Liu Y, Zhang B, Economic strategies in the issue of controlling AIDS, Proceeding of IEEE/ICME International Conference, 2007, 1601- 1608

20. Shmueli A, Israelis evaluate their health care system before and after the introduction of the national health insurance law, Health Policy, 2003(63), 279-287

21. Kiil A, What characterises the privately insured in universal health care systems? A review of the empirical evidence, Health Policy, 2012(106), 60-75

22. Wensing M, Baker R, Szecsenyi J, Grol R, On behalf of the EUROPEP Group, Impact of national health care systems on patient evaluations of general practice in Europe, Health Policy, 2004(68), 353-357

23. Gu Xing-Yuan, Tang Sheng-Lan, Reform of the Chinese health care financing system, Health Policy, 1995(32), 181-191

24. Yaesoubi R, Roberts SD, Payment contracts in a preventive health care system: A perspective from Operations Management, Journal of Health Economics, 2011(30), 1188-1196

25. Avgerinos ED, Koupidis SA, Filippou DK, Impact of the European Union enlargement on health professionals and health care systems, Health Policy, 2004(69), 403-408

26. Zu H, Wang Q, Dong M, Ma L, Yin L, Yang Y, Compressed Sensing Based Fixed-Point DCT Image Encoding, Advances in Computational Mathematics and its Applications, 2012(2), 237-240

27. Wang Q, Li M, Xia LC, Wen G, Zu H, Gao M(2013), Genetic Analysis about Differentiation of Helper T Lymphocytes, Genetics and Molecular Research, in press

28. Xia L, Zhou C, Phase transition in sequence unique reconstruction, Journal of Systems Science and Complexity 2007(20), 18-29

29. Zhang SW, Li YJ, Xia L, Pan Q, PPLook: an automated data mining tool for protein-protein interaction, BMC bioinformatics 2011(11), 326

30. He PA, Xia L, Oligonucleotide profiling for discriminating bacteria in bacterial communities, Combinatorial Chemistry & High Throughput Screening 2007(10), 247-255

31. Steele JA, Countway PD, Xia L, et al., Marine bacterial, archaeal and protistan association networks reveal ecological linkages, The ISME Journal 2011(5), 1414-1425.

Corresponding Author
Qixin Wang,
Department of Mathematics,
University of Southern California
Los Angeles,
United States of America,
E-mail: qixin.wang@usc.edu